\documentclass[10pt,twocolumn]{revtex4}

\usepackage[utf8]{inputenc}
\usepackage{hyperref}
\usepackage{epsfig}
\usepackage{graphicx}
\usepackage{amsmath}
\usepackage{amssymb}
\usepackage{color}

\graphicspath{{Images/}}

\begin{document}

\title{
Short-time Fourier transform laser Doppler holography
}
\author{Benjamin Samson}
\author{Michael Atlan}

\affiliation{
Institut Langevin. Centre National de la Recherche Scientifique (CNRS) UMR 7587, Institut National de la Sant\'e et de la Recherche M\'edicale (INSERM) U 979, Universit\'e Pierre et Marie Curie (UPMC), Universit\'e Paris Diderot. \'Ecole Sup\'erieure de Physique et de Chimie Industrielles - 1 rue Jussieu. 75005 Paris. France
}

\date{\today}
\begin{abstract}
We report a demonstration of laser Doppler holography at a sustained acquisition rate of 250 Hz on a 1 Megapixel complementary metal–oxide–semiconductor (CMOS) sensor array and image display at 10 Hz frame rate. The holograms are optically acquired in off-axis configuration, with a frequency-shifted reference beam. Wide-field imaging of optical fluctuations in a 250 Hz frequency band is achieved by turning time-domain samplings to the dual domain via short-time temporal Fourier transformation. The measurement band can be positioned freely within the low radio-frequency (RF) spectrum by tuning the frequency of the reference beam in real-time. Video-rate image rendering is achieved by streamline image processing with commodity computer graphics hardware. This experimental scheme is validated by a non-contact vibrometry experiment.
\end{abstract}

\maketitle

Though effective for single-point analysis \cite{Castellini2006}, laser Doppler measurements are more difficult to perform in wide-field imaging configuration, because of a technological challenge : digital image frames have to be read out at kHz rates and beyond to perform short-time discrete Fourier transforms (DFT) \cite{AllenRabiner1977}. Recently, image-plane laser Doppler recordings with a high throughput CMOS camera in conjunction with short-time DFT calculations by a field programmable gate array (FPGA) reportedly enabled continuous monitoring of blood perfusion in the mm/s range. Full-field flow maps of 480 $\times$ 480 pixels were rendered at a rate of 14 Hz, obtained from image recordings at a frame rate of 14.9 kHz \cite{Leutenegger2011}. For transient dynamics imaging of faster phenomena, high throughput laser Doppler schemes were designed by multipoint \cite{Kilpatrick2008} or time multiplexing \cite{FuGuo2011} approaches. High speed holography enabled offline vibrometry from time-resolved optical phase measurements \cite{Pedrini2006}. Heterodyne holography, as a variant of time-averaged holography \cite{Powell1965, PicartLeval2003} with a strobe \cite{Hariharan1987} or a frequency-shifted reference beam \cite{Aleksoff1971, JoudVerpillat2009}, is appropriate for steady-state (at the scale of the exposure time) mechanical vibrations mapping. Advances in reconstruction techniques of optically-measured digital holograms with Graphics Processing Units (GPUs) \cite{ShimobabaSato2008, Ahrenberg2009} have led to real-time holographic screening of a single vibration frequency, demonstrated in this regime \cite{SamsonVerpillat2011}.\\

In this letter, we report an experimental demonstration of video-rate image reconstruction and display of laser fluctuation spectra from high speed holographic measurements. Sustained Fresnel reconstruction of off-axis holograms at 250 Hz and 10 Hz rendering by short-time DFTs is performed. Images and RF spectra of a thin metal plate's out-of-plane vibration modes around 3.2 kHz are presented.\\

\begin{figure}[]
\centering
\includegraphics[width = 8.3 cm]{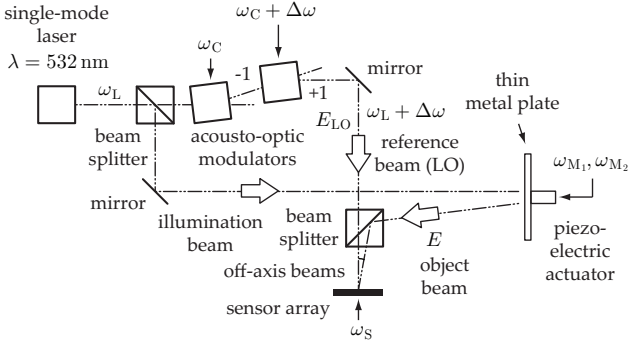}
\caption{Optical arrangement. The main laser beam is split into two channels, forming a Mach-Zehnder interferometer. In the object channel, the optical field $E$ is backscattered by a metal plate in vibration is phase-modulated according to eq. \eqref{eq_E_defn}. In the reference channel, the optical field $E_{\rm LO}$ is frequency-shifted by two acousto-optic modulators from which alternate diffraction orders ($\pm1$) are selected, yielding an optical LO of the form of eq. \eqref{eq_ELO_defn}. Images of the metal plate are computed numerically from the holographic measurement of the diffracted object beam beating against the frequency-shifted local oscillator beam, with standard image rendering algorithms \cite{VerrierAtlan2011}.}
\label{fig_Setup}
\end{figure}

\begin{figure}[]
\centering
\includegraphics[width = 8.3 cm]{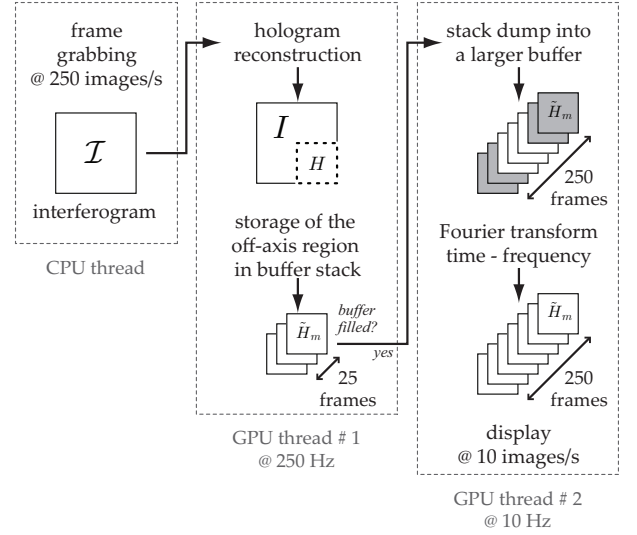}
\caption{Algorithmic layout of holographic rendering. Raw interferograms are recorded by the main CPU thread. Spatial Fresnel transforms are performed by the first GPU thread. Short-time temporal Fourier transforms are performed by the second GPU thread.}
\label{fig_algo}
\end{figure}

\begin{figure}[]
\centering
\includegraphics[width = 8.3 cm]{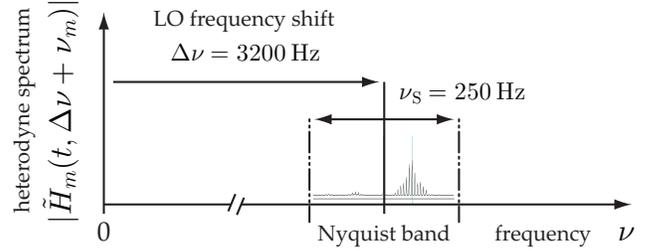}
\caption{Sketch of the spectral screening region.}
\label{fig_SpectralRegions}
\end{figure}

The optical setup, sketched in fig.\ref{fig_Setup}, is similar to the one reported in the demonstration of video-rate vibrometry at a single frequency \cite{SamsonVerpillat2011}, at a difference that a high throughput CMOS camera is used to achieve megapixel recordings at 250 frames per second. An off-axis, frequency-shifted Mach-Zehnder interferometer is used to perform a multipixel heterodyne detection of an object field $E$ beating against a separate local oscillator (LO) field $E_{\rm LO}$, in reflective geometry. The main optical radiation field is provided by a 100 mW, single-mode laser (wavelength $\lambda = 532$ nm, optical frequency $\nu_{\rm L} = \omega_{\rm L}/(2 \pi) = 5.6 \times 10^{14} \, \rm Hz$, Oxxius SLIM 532). The optical frequency of the LO beam is shifted by an arbitrary quantity $\Delta \nu$ in the low RF range by two acousto-optic modulators (AA-electronics, MT80-A1.5-VIS). The object studied is a thin metal plate with hexagonal holes, shined over $\sim 30 \, {\rm mm} \times 30 \, {\rm mm}$ with $\sim$ 50 mW of impinging light. It is excited with a piezo-electric actuator (PZT, Thorlabs AE0505D08F), vibrating sinusoidally, driven at 10 V. The structure's vibrations provoke a local phase modulation $\phi$ (eq.\ref{eq_phi_defn}) of the backscattered optical field $E$. Interference patterns are measured with a Basler A504k camera (Micron MV13 progressive scan CMOS sensor array of $1280 \times 1024$ pixels, quantum efficiency $\sim$ 25 \% at 532 nm). The camera is run in external trigger mode at $\nu_{\rm S} = \omega_{\rm S} / (2 \pi) = 250 \, \rm Hz$, at 8 bit/pixel quantization. Images of the central $1024 \times 1024$ pixels region are recorded. The image acquisition is interfaced with a National Instruments NI PCIe-1433 frame grabber. Each raw interferogram digitally acquired at time $t$, noted ${\cal I}(t) =  \left| E(t) + E_{\rm LO}(t) \right| ^2$ is dumped to a $1024 \times 1024 \times 1 \, \rm byte$ frame buffer in the GPU RAM of a NVidia GTX 580 graphics card by a CPU thread (fig. \ref{fig_algo}). The object field of complex amplitude ${\cal E}$ is noted
\begin{equation}
E = {\cal E} \exp \left( i \omega _{{\rm L}} t + i \phi(t) \right) 
\label{eq_E_defn}
\end{equation}
where $\omega _{\rm L} = 2 \pi \nu _{\rm L}$  and $\phi(t)$ is the fluctuating phase, as a result of optical path length modulation. The acousto-optic modulators enable the optical LO field of complex amplitude $\cal E_{\rm LO}$ to be detuned by $\Delta \nu = \Delta \omega / (2 \pi)$
\begin{equation}
E_{\rm LO} = {\cal E}_{\rm LO} \exp \left( i \omega_{\rm L} t + i \Delta \omega t \right).
\label{eq_ELO_defn}
\end{equation}
Holographic image rendering from each recorded interferogram is performed with a numerical Fresnel transform. The hologram $I$, back-propagated to the object plane, is calculated by forming the Fast Fourier Transform (FFT) ${\cal F}$ of the product of ${\cal I}$ with a quadratic phase map, depending on the relative curvature of the wavefronts of $E$ and $E_{\rm LO}$ in the sensor plane \cite{VerrierAtlan2011}. This calculation is handled by the GPU (thread \#1, Fig. \ref{fig_algo}), by an algorithm elaborated with Microsoft Visual C++ 2008 and NVIDIA's Compute Unified Device Architecture (CUDA) software development kit 3.2, on single precision floating point arrays. The practical implementation of free-space propagation with a discrete Fresnel transform \cite{PicartLeval2003} yields complex-valued holograms carried by the cross-terms of the interference pattern $I = |E|^2 + |E_{\rm LO}|^2 + E^* E_{\rm LO} + E E^*_{\rm LO}$ reconstructed in the object plane. In off-axis configuration \cite{Cuche2000}, the zero-order terms $|E|^2$ and $|E_{\rm LO}|^2$ and the twin-image term $E^* E_{\rm LO}$ can be filtered-out. After filtering, the remaining complex-valued contribution to the off-axis hologram is 
\begin{equation}\label{eq_interferogram}
H(t) = E E^*_{\rm LO} = {\cal E} {\cal E}^*_{\rm LO} {\exp}{\left( i \phi(t) - i \Delta \omega t \right)}.
\end{equation}
The heterodyne spectrum of the radiation field $E$ is detected by a short-time discrete Fourier transform (DFT) of $H(t)$ over $N = 250$ consecutive samples (fig. \ref{fig_algo}, thread \# 2), 10 times per second. The $m$-th Fourier component of the DFT,
\begin{equation}
     \tilde{H}_m (t) = \sum_{n=1}^{N} H \left( t - n / \nu_{\rm S} \right) \exp \left(- 2 i \pi m n / N \right)
    \label{Eq_I_tilde}
\end{equation}
is a heterodyne measurement of the laser fluctuation spectrum at time $t$, at frequency $\Delta \nu + \nu_m$
\begin{equation}
\tilde{H}_m (t) = \tilde{H}(t,\Delta \nu + \nu_m).
\label{Eq_I_tilde_nu_t}
\end{equation}
The discrete frequencies $\nu_m$ of the measured spectra lie within the Nyquist limits of the camera bandwidth $\pm \nu_{\rm S}/2$, while the LO detuning frequency $\Delta \nu$ can be set arbitrarily by the acousto-optic modulators (fig. \ref{fig_SpectralRegions}).\\

\begin{figure}[]
\centering
\includegraphics[width = 8.3 cm]{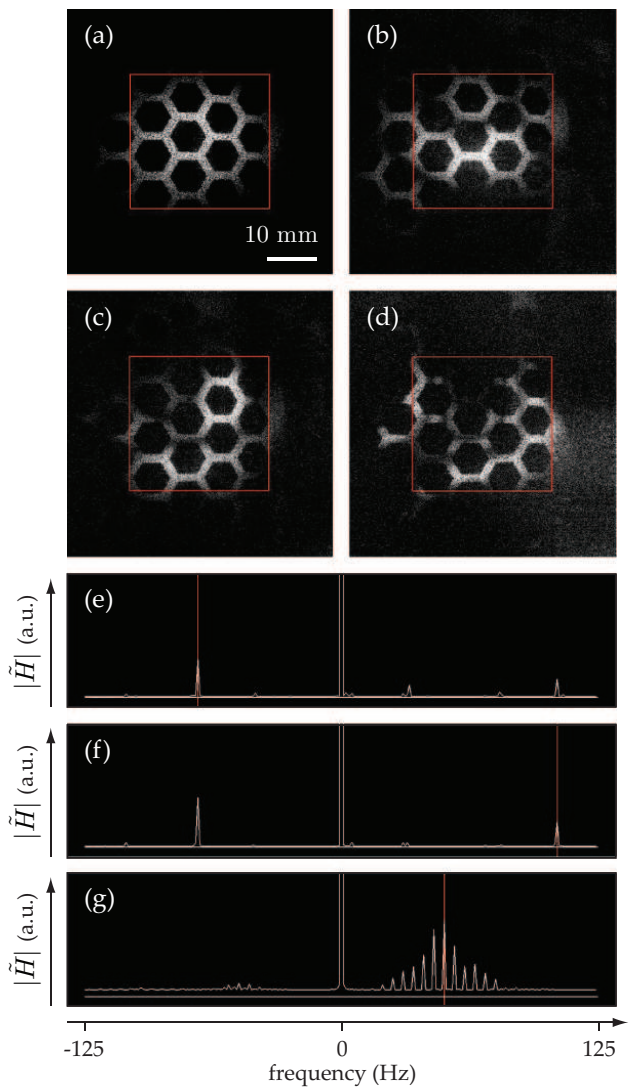}
\caption{Short-time Fourier transform vibration maps and spectra of a metal structure excited sinusoidally. No excitation (a). Excitation at $P=2$ frequencies : $\nu_{{\rm M}_1} = 3200 + 100 \, \rm Hz$ and $\nu_{{\rm M}_2} = 3200 - 70 \, \rm Hz$ (b-c,e-f). Excitation at $P=2$ frequencies : $\nu_{{\rm M}_1} = 3200 + 50 \, \rm Hz$ and $\nu_{{\rm M}_2} = 3200 + 45 \, \rm Hz$ (d,g). LO detuning : $\Delta \nu = 3200 \, \rm Hz$; The movie of the experiment is reported in media 1.}
\label{fig_maps}
\end{figure}
We assessed the thin metal plate's out-of-plane vibration modes around 3.2 kHz with the presented holographic approach. The metallic structure was excited sinusoidally at one ($P=1$) or two ($P=2$) frequencies $\nu _{{\rm M}_1}$ and $\nu _{{\rm M}_2}$. In either case, the resulting out-of-plane motion at a given point of the surface of the plate, considered as a linear medium for acoustic waves, is
\begin{equation}
z(t) = \sum _{p=1} ^{P} z_p \sin \left( \omega _{{\rm M}_p} t \right)
\label{eq_z_defn}
\end{equation}
where $\omega _{{\rm M}_p} = 2 \pi \nu _{{\rm M}_p}$ and $z_p$ are the angular frequency and the local amplitude of each component, respectively. The phase modulation of the backscattered light is
\begin{equation}
\phi(t) = \frac{4 \pi}{\lambda} z(t) = \sum _{p=1} ^{P} \phi_p \sin \left( \omega _{{\rm M}_p} t \right)
\label{eq_phi_defn}
\end{equation}
where $\phi_p = 4 \pi z_p / \lambda$. The temporal part of the field undergoing sinusoidal phase modulation can be decomposed in a basis of Bessel functions with the Jacobi–Anger identity. If $P=1$, the object field is
\begin{equation}
E = {\cal E} \exp \left( i \omega _{{\rm L}} t \right) \sum_{n=-\infty}^{\infty} J_n \left( \phi_1 \right) \exp \left( i n \omega_{{\rm M}_1} t \right)
\label{eq_E_P1}
\end{equation}
where $J_n$ is the Bessel function of the first kind of rank $n$. The only remaining low frequency, camera-filtered term in eq. \ref{eq_interferogram} beating around $\Delta \nu$, within the Nyquist domain (i.e. for $ | \nu_{{\rm M}_1} - \Delta \nu | < \nu_{\rm S}/2 $) is
\begin{equation}
H_{\rm LF}(t) = {\cal E} _{\rm LO} ^* {\cal E} J_1(\phi_1) \exp \left( i \omega_{{\rm M}_1} t - i \Delta \omega t \right).
\label{eq_EEprime_P1}
\end{equation}
This modulated hologram yields a single component in the short-time DFT spectrum $\tilde{H}_m (t)$, at the frequency $\nu_m = \nu_{{\rm M}_1} - \Delta \nu$. In the first part of the movie (media 1, when $P=1$), the excitation frequency $\nu_{{\rm M}_1}$ was swept from 3210 Hz to 3290 Hz, and the LO was detuned by $\Delta \nu = 3200 \, \rm Hz$; the measurement frequency $\nu_m$ of the short-time DFT was swept concurrently from 10 Hz to 90 Hz, in 5 Hz steps. The reported spectra result from the magnitude $|\tilde{H}(t,\Delta \nu + \nu_m)|$ averaged within the red square superimposed on the vibration maps.\\

The metallic structure was then excited sinusoidally at two frequencies : $\nu _{{\rm M}_1}$, and $\nu _{{\rm M}_2}$. The object field undergoing phase modulation from a double excitation ($P=2$ in eq. \ref{eq_phi_defn}) takes the form 
\begin{equation}
E = {\cal E} \exp \left( i \omega _{{\rm L}} t \right) 
\prod_{p=1}^{2} \, \sum_{n=-\infty}^{\infty} J_n \left( \phi_p \right) \exp \left( i n \omega_{{\rm M}_p} t \right).
\label{eq_E_prod_sum}
\end{equation}
The terms of eq. \ref{eq_interferogram} modulated at frequencies within the camera bandwidth $\pm \omega_{\rm S} / 2$ are actually measured. The others are filtered out. The temporal part of the camera-filtered hologram reduces to the low frequency component
\begin{eqnarray}
H_{\rm LF}(t) = {\cal E} _{\rm LO} ^* {\cal E} {\rm e}^{ -i \Delta \omega t } \sum_{n =- \infty}^{\infty} c_{n,1} c_{-n+1,2}
\label{eq_EEprime_P2}
\end{eqnarray}
where 
\begin{equation}
c _{n,p} = J_n(\phi_p) \exp \left( i n \omega_{{\rm M}_p} t \right).
\label{eq_cnp}
\end{equation}
Eq.\ref{eq_EEprime_P2} yields the frequency comb observed in fig.\ref{fig_maps}(g), whose peaks are separated by $|\nu_{{\rm M}_2} - \nu_{{\rm M}_1}| = 5 \, \rm Hz$. In the second part of the movie reported in media 1, the first excitation frequency was set to $\nu_{{\rm M}_1} = 3290 \, \rm Hz$, the second one was swept from $\nu_{{\rm M}_2} = 3290 \, \rm Hz$ to $\nu_{{\rm M}_2} = 3250 \, \rm Hz$, in 5 Hz steps. The frequency comb broadened with $|\nu_{{\rm M}_2} - \nu_{{\rm M}_1}|$ in accordance with equations \ref{eq_EEprime_P2} and \ref{eq_cnp}. For a larger frequency difference $|\nu_{{\rm M}_2} - \nu_{{\rm M}_1}|$, only two lines of the comb are visible (fig.\ref{fig_maps}(e,f)).\\

In conclusion, we performed laser Doppler imaging from sustained sampling of 1 Mega pixel interferograms at a throughput of 250 Mega bytes per second, and rendering of 0.25 Mega pixel off-axis heterodyne holograms by short-time discrete Fourier transform with a refreshment rate of 10 Hz. This demonstration was made with commodity computer graphics hardware. We reported video-rate optical monitoring of out-of-plane vibration amplitudes in a frequency band of 250 Hz, shifted by 3.2 kHz from DC. This demonstration opens the way to high bandwidth laser Doppler holography in real time.\\

We thank Francois Bruno and Fiona Quinlan-Pluck careful rereading of the manuscript. We gratefully acknowledge support from Fondation Pierre-Gilles de Gennes (FPGG014), Agence Nationale de la Recherche (ANR-09-JCJC-0113, ANR-11-EMMA-046), r\'egion \^Ile-de-France (C'Nano, AIMA), and the "investments for the future" program (LabEx WIFI: ANR-10-IDEX-0001-02 PSL*).\\

\bibliographystyle{unsrt}%

\end{document}